\documentclass[aps,prb,superscriptaddress,twocolumn,showpacs,preprintnumbers,amsmath,amssymb]{revtex4}

\usepackage[dvips]{graphicx,color}
\usepackage{dcolumn}
\usepackage{bm}
\usepackage{ulem}
\usepackage{soul}
\begin{document}

\title{Spin treacle in a frustrated magnet observed with spin current}

\author{Hiroki~Taniguchi}
\affiliation{Department of Physics, Graduate School of Science, Osaka University, Toyonaka 560-0043, Japan}
\author{Mori~Watanabe}
\affiliation{Department of Physics, Graduate School of Science, Osaka University, Toyonaka 560-0043, Japan}
\author{Takashi~Ibe}
\affiliation{Department of Physics, Graduate School of Science, Osaka University, Toyonaka 560-0043, Japan}
\author{Masashi~Tokuda}
\affiliation{Department of Physics, Graduate School of Science, Osaka University, Toyonaka 560-0043, Japan}
\author{Tomonori~Arakawa}
\affiliation{Department of Physics, Graduate School of Science, Osaka University, Toyonaka 560-0043, Japan}
\affiliation{Center for Spintronics Research Network, Osaka University, Toyonaka 560-8531, Japan}
\author{Toshifumi~Taniguchi}
\affiliation{Department of Earth and Space Science, Graduate School of Science, Osaka University, Toyonaka 560-0043, Japan}
\author{Bo~Gu}
\affiliation{Kavli Institute for Theoretical Sciences, and CAS Center for Excellence in Topological Quantum Computation, University of Chinese Academy of Sciences, Beijng 100190, China}
\affiliation{Physical Science Laboratory, Huairou National Comprehensive Science Center, Beijing 101400, China}
\author{Timothy~Ziman}
\affiliation{Institut Laue-Langevin, BP 156, 41 Avenue des Martyrs, 38042 Grenoble Cedex 9, France}
\affiliation{Universit\'e Grenoble Alpes, Centre National de la Recherche Scientifique, LPMMC, 38000 Grenoble, France}
\affiliation{Kavli Institute for Theoretical Sciences, and CAS Center for Excellence in Topological Quantum Computation, University of Chinese Academy of Sciences, Beijng 100190, China}
\author{Sadamichi~Maekawa}
\affiliation{Center for Emergent Matter Science, RIKEN, Wako 351-0198, Japan}
\affiliation{Kavli Institute for Theoretical Sciences, and CAS Center for Excellence in Topological Quantum Computation, University of Chinese Academy of Sciences, Beijng 100190, China}
\author{Kensuke~Kobayashi}
\affiliation{Department of Physics, Graduate School of Science, Osaka University, Toyonaka 560-0043, Japan}
\affiliation{Institute for Physics of Intelligence and Department of Physics, Graduate School of Science, The University of Tokyo, Tokyo 113-0033, Japan}
\author{Yasuhiro~Niimi}
\email{niimi@phys.sci.osaka-u.ac.jp}
\affiliation{Department of Physics, Graduate School of Science, Osaka University, Toyonaka 560-0043, Japan}
\affiliation{Center for Spintronics Research Network, Osaka University, Toyonaka 560-8531, Japan}

\date{\today}
\pacs{72.25.Ba, 72.25.Mk, 75.70.Cn, 75.75.-c}

\begin{abstract}
By means of spin current, the flow of spin angular momentum, 
we find a regime of ``spin treacle'' in a frustrated magnetic system. 
To establish its existence, we have performed spin transport measurements 
in nanometer-scale spin glasses. 
At temperatures high enough that the magnetic moments 
fluctuate at high frequencies, the spin Hall angle, the conversion yield between spin current 
and charge current, is independent of temperature. 
The spin Hall angle starts to decrease at a certain temperature $T^{*}$ and 
completely vanishes at a lower temperature. 
We argue that the latter corresponds to the spin freezing temperature $T_{\rm f}$ of the
nanometer-scale spin glass, where the direction of conduction electron spin is 
randomized by the exchange coupling with the localized moments. 
The present experiment \textit{quantitatively} verifies the existence of 
a distinct ``spin treacle" between $T_{\rm f}$ and $T^{*}$. 
We have also quantified a time scale of fluctuation of local magnetic moments
in the spin treacle 
from the spin relaxation time of conduction electrons.
\end{abstract}
\maketitle

\section{Introduction}

For several decades, spin glass (SG) has been extensively studied as a prototype of 
complex system characterized 
by frustration and randomness~\cite{Binder_Young_1986, Kawamura_Taniguchi_2015, Young_1997}. 
Thus, understanding its ground state and any excitation modes is of importance not only 
in condensed matter physics but also in a wide range of scientific areas~\cite{nature_1999}. 
Recently, it has drawn renewed interest from the viewpoint of quantum information engineering, 
where it is well-known as the basis of quantum annealing~\cite{Kadowaki_Nishimori_1998,Johnson_2011}.

SG appears when magnetic impurities are randomly distributed 
in a nonmagnetic noble metal host~\cite{Cannella_PRB_1972,Mydosh_1993}. 
The interaction between the localized moments is mediated by conduction electron spins, 
which is referred to as the Ruderman-Kittel-Kasuya-Yosida (RKKY) 
interaction~\cite{Ruderman_Kittel_1954,Kasuya_1956,Yosida_1957}. 
As a result of the random distribution of magnetic impurities and the RKKY interaction, 
SG exhibits a cusp anomaly in the magnetic susceptibility as a function of temperature 
under zero field cooling (ZFC), and takes a constant value under 
filed cooling (FC)~\cite{Nagata_1979}. 
The cusp temperature is called spin freezing temperature $T_{\rm f}$, 
below which the magnetic moments freeze randomly. 

According to recent quantum coherence~\cite{Capron_2013} and 
spin transport measurements~\cite{Niimi_PRL_2015} in nanometer-scale SG devices, 
the magnetic moments fluctuate at higher temperatures than $T_{\rm f}$ and 
keep fluctuating even below $T_{\rm f}$. Here we note that $T_{\rm f}$ in 
Refs.~\onlinecite{Capron_2013} and \onlinecite{Niimi_PRL_2015} was determined 
from magnetization measurements with the thin films. 
However, this $T_{\rm f}$ might be different from that of nanoscale wires. 
It is in general very difficult to detect tiny magnetic signals in nanoscale samples 
on-chip with the conventional methods such as 
magnetization~\cite{Cannella_PRB_1972,Mydosh_1993,Nagata_1979}, 
electron spin resonance (ESR)~\cite{date_jpsj_1969,chien_jap_1994},
muon spin resonance ($\mu$SR)~\cite{Uemura_PRB_1985,Campbell_PRL_1994}, 
nuclear magnetic resonance~\cite{MacLaughlin_PRL_1976,Bloyet_PRL_1978,Alloul_PRL_1979}, 
and neutron scattering measurements~\cite{murani_prl_1978,mezei_jmmm_1979}.
Thus, it is an important and challenging task to develop a new experimental method 
to characterize $T_{\rm f}$ for nanoscale samples.

\begin{figure}
\begin{center}
\includegraphics[width=5cm]{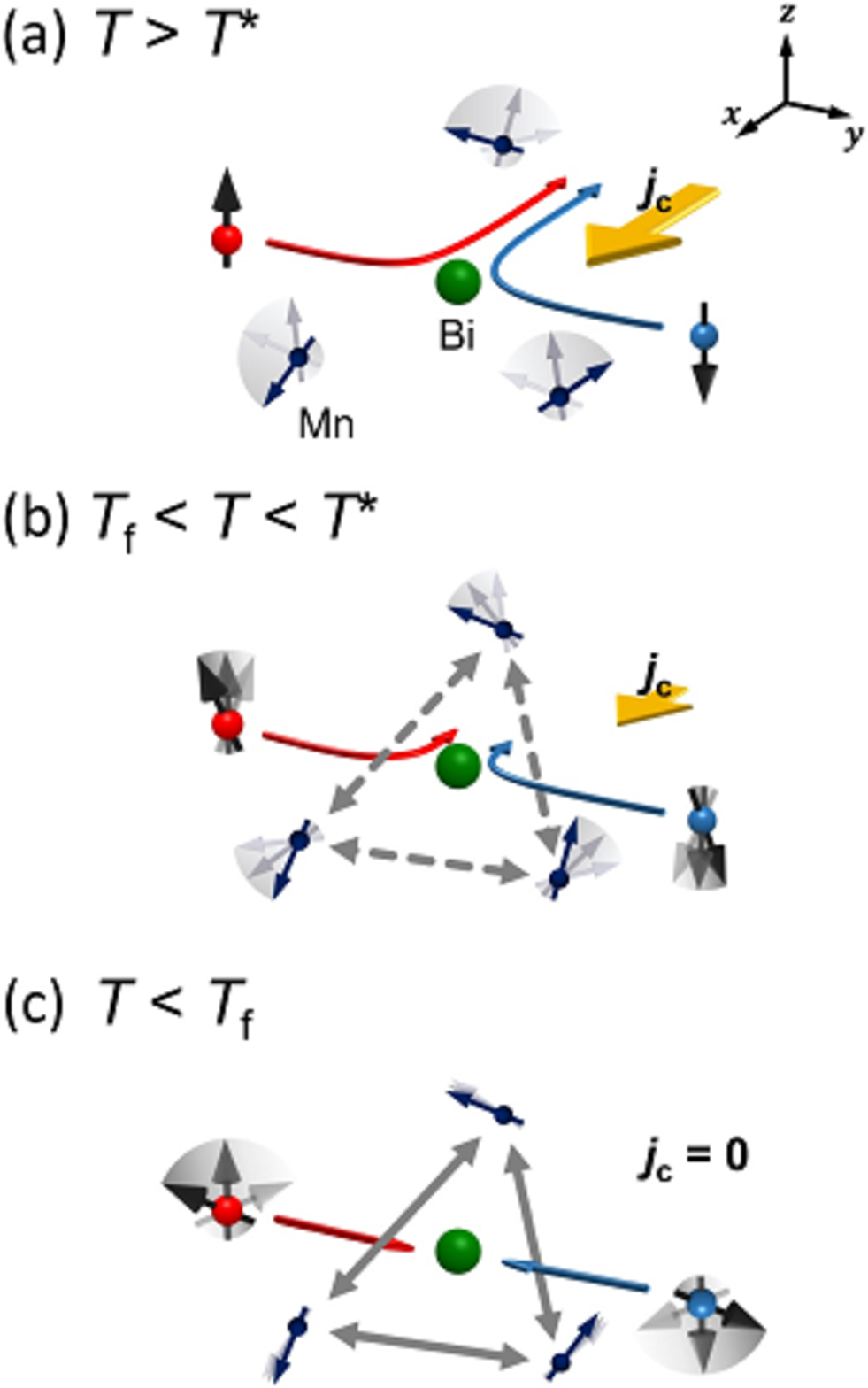}
\caption{Illustrations of ISHE in CuMnBi at three different temperature regimes: 
(a) $T > T^{*}$ (PM), 
(b) $T_{\rm f} < T < T^{*}$ (ST), and 
(c) $T < T_{\rm f}$ (SG). 
Black arrows with red and blue spheres are 
conduction electrons with spin-up and spin-down, respectively, 
and red and blue arrows show those trajectories. 
The shadows indicate fluctuations of conduction electron spins and magnetic moments of Mn. 
Yellow and gray arrows indicate the charge current density ${\bm j}_{\rm c}$ generated at the Bi site 
(green sphere) due to the ISHE and a magnetic interaction between the Mn sites, respectively. 
The $x$, $y$, and $z$ axes in (a) correspond to the directions of ${\bm j}_{\rm c}$, ${\bm j}_{\rm s}$, 
and ${\boldsymbol \sigma}$ above $T^{*}$, respectively.}
\label{fig:figure1}
\end{center}
\end{figure}

In this paper, we propose a new method to determine $T_{\rm f}$ of 
nanoscale SG using spin current, flow of spin angular momentum only. 
Some of the present authors have already shown that the inverse spin Hall effect (ISHE) 
which enables us to convert the spin current into the charge current starts to decrease 
at $T^{*}$, 4-5 times higher temperature than $T_{\rm f}$ of the thin film~\cite{Niimi_PRL_2015}. 
Here we find a temperature region where the ISHE signal vanishes completely. 
This corresponds to the SG state in nanowires, which was 
difficult to evaluate quantitatively by previous theories and experiments.  
Thanks to the new definition of $T_{\rm f}$ for SG nanowires proposed in the present work, 
we are able to verify that there is an additional region ($T_{\rm f} < T < T^{*}$)
in between the SG state and the paramagnetic (PM) state, that 
we call \textit{spin treacle} (ST), as highlighted in Fig.~\ref{fig:figure1} 
(we will explain this figure in more detail in Sec. III~B). 
These experimental results show that the ISHE is a powerful method to 
detect spin dynamics and spin fluctuations even in nanoscale samples 
with complex magnetic structures on-chip.
In addition, we have measured the correlation time $\tau_{\rm c}$ 
of the localized moments in ST through the spin relaxation time of conduction electrons.

\section{Experimental details}

The SHE device is based on a lateral spin valve structure 
where a SG (Cu$_{99.5-x}$Mn$_{x}$Bi$_{0.5}$ or Au$_{78}$Fe$_{22}$) nanowire is inserted 
in between two ferromagnetic permalloy (${\rm Ni}_{81}{\rm Fe}_{19}$, hereafter Py) wires, and 
those three nanowires are bridged by a Cu wire. Samples were patterned using 
electron beam lithography onto a thermally oxidized silicon 
substrate coated with polymethyl-methacrylate (PMMA) resist 
for depositions of Py and Cu, or coated with ZEP 520A resist 
for depositions of CuMnBi and AuFe. 

A pair of Py wires was first deposited using an electron 
beam evaporator under a base pressure of $6\times10^{-7}$ Pa. 
The width and thickness of the Py wires are 100 and 30 nm, 
respectively.
The CuMnBi (or AuFe) middle wire was next deposited 
by magnetron sputtering using a CuMnBi (or AuFe) target. 
The Bi concentrations used in this work 
was fixed at 0.5\%, while 
the Mn concentration $x$ was changed from 4.2 to 10.6\% for Cu$_{99.5-x}$Mn$_{x}$Bi$_{0.5}$ 
(i.e., Cu$_{95.3}$Mn$_{4.2}$Bi$_{0.5}$, 
Cu$_{91.3}$Mn$_{8.2}$Bi$_{0.5}$, and Cu$_{88.9}$Mn$_{10.6}$Bi$_{0.5}$). 
The width and thickness of the CuMnBi middle wire 
are 250 and 20~nm, respectively. 
For the AuFe middle wire, on the other hand, the width and the thickness are 
120~nm and 30~nm, respectively. 
The post-baking temperature for the PMMA resist was kept 
below 100$^{\circ}$C after the deposition of CuMnBi alloys 
since bismuth has a low melting temperature (270$^{\circ}$C). 
Before deposition of a Cu bridge, we performed a careful Ar 
ion beam etching for 1 minute in 
order to clean the surfaces of the Py and CuMnBi middle wires. 
After the Ar ion etching, the device was moved to another 
chamber without breaking a vacuum and subsequently the Cu 
bridge was deposited by a Joule heating evaporator using a 
99.9999\% purity source. 
Both the width and thickness of the Cu bridge are 100~nm. 

The SHE and nonlocal spin valve (NLSV) measurements have been carried out 
using an \textit{ac} lock-in amplifier and a $^{4}$He flow cryostat. 
The magnetization measurements were performed 
with a commercial superconducting quantum interferometry
device (SQUID) magnetometer, MPMS (Quantum Design)~\cite{taniguchi_prl_2004}.

\section{Experimental results}

\subsection{Magnetization measurements of bulk and thin film of CuMnBi}

\begin{figure}[htbp]
\begin{center}
\includegraphics[width=8.5cm]{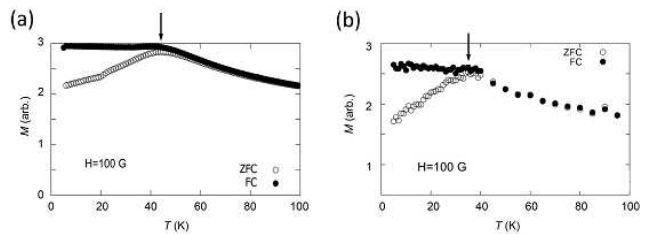}
\caption{Magnetizations $M$ of (a) bulk Cu$_{88.9}$Mn$_{10.6}$Bi$_{0.5}$ and 
(b) 140~nm thick Cu$_{88.9}$Mn$_{10.6}$Bi$_{0.5}$ film. 
The open and closed circles indicate $M$ under ZFC and FC, respectively. 
The arrows in (a) and (b) indicate $T_{\rm f}$ of bulk and film samples, respectively.}
\label{fig:figure2}
\end{center}
\end{figure}

To determine $T_{\rm f}$ of bulk CuMnBi and thin CuMnBi film, 
we performed the magnetization $M$ measurements. 
Figures~\ref{fig:figure2}(a) and \ref{fig:figure2}(b) show $M$ of bulk Cu$_{88.9}$Mn$_{10.6}$Bi$_{0.5}$ and 
140~nm thick Cu$_{88.9}$Mn$_{10.6}$Bi$_{0.5}$ film. 
A clear cusp structure can be seen both in bulk and thin film samples, but 
$T_{\rm f}$ of thin film is 80-90\% of that of bulk. 
We observe the similar tendency for the other Mn concentrations and also for an AuFe alloy
(see Supplemental Material in Ref.~\onlinecite{supplemental_material}). 

\subsection{ISHE measurements in CuMnBi nanowire}

\begin{figure}
\begin{center}
\includegraphics[width=8.5cm]{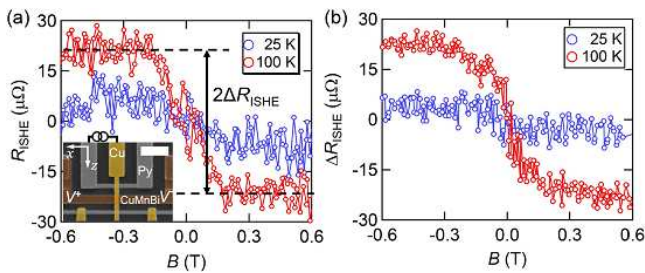}
\caption{ISHE resistances ($R_{\rm ISHE} \equiv V/I$) of ${\rm Cu}_{99.5-x}{\rm Mn}_{x}{\rm Bi}_{0.5}$ 
measured at typical temperatures ((a) $x = 4.2$, (b) $x = 10.6$). The amplitude of the ISHE resistance 
$\Delta R_{\rm ISHE}$ is defined in (a). The inset of (a) shows a scanning electron micrograph of 
typical spin Hall device and a schematic of the ISHE measurement circuit. 
The white bar in the inset corresponds to ${\rm 1\ \mu m}$. 
The $x$ and $z$ axes in (a) are the same as those in Fig.~\ref{fig:figure1}(a).}
\label{fig:figure3}
\end{center}
\end{figure}

After characterizing $T_{\rm f}$ of bulk and thin film of CuMnBi, 
we now focus on the spin transport property obtained with CuMnBi nanowires. 
The ISHE measurements were performed using the spin absorption method 
in the lateral spin valve structure~\cite{Niimi_rpp_2015}. 
The inset of Fig.~\ref{fig:figure3}(a) shows a typical spin Hall device. 
As explained in the previous section, the SHE device consists of 
two Py wires and ${\rm Cu}_{99.5-x}{\rm Mn}_{x}{\rm Bi}_{0.5}$ 
($x = 4.2, 8.2,$ and $10.6$) wire, 
which are bridged by a Cu stripe. 
CuMn is a typical SG material~\cite{Nagata_1979}, 
but a small amount ($0.5\%$) of Bi impurities is added to CuMn, in order to induce a 
large ISHE~\cite{fert_1981,fert_levy_2011,niimi_2012,niimi_prb_2014}. 
By flowing an electric current $I$ from 
one of the Py wires [top wire in the inset of Fig.~\ref{fig:figure3}(a)] 
to the Cu stripe, spin accumulation is generated at the interface between Py and Cu. 
A spin current flows in the Cu stripe (downward) as a result of diffusion process 
of the spin accumulation. When a strong spin-orbit material, i.e., CuMnBi, 
is placed underneath the Cu stripe within the spin diffusion length of Cu 
($\sim1~{\rm \mu m}$ at low temperatures), a part of spin current is injected into 
CuMnBi because of a stronger spin-orbit interaction 
due to the Bi impurities~\cite{Niimi_rpp_2015}. When the magnetization of the Py wire 
is fully polarized along the Cu stripe ($|B| > 0.3$~T), 
a spin-to-charge conversion is generated along the 
CuMnBi wire direction as shown in Fig.~\ref{fig:figure1}(a):
\begin{eqnarray}
{\bm j}_{\rm c} = \alpha_{\rm H}^{\rm CuMnBi} ({\bm j}_{\rm s}\times {\boldsymbol \sigma}) 
= \alpha_{\rm H}^{\rm CuMnBi} j_{\rm s} \hat{\mbox{\boldmath $x$}}, \label{eq1}
\end{eqnarray}
where ${\bm j}_{\rm c}$, $\alpha_{\rm H}^{\rm CuMnBi}$, ${\bm j}_{\rm s}$, ${\boldsymbol \sigma}$, 
and $\hat{\mbox{\boldmath $x$}}$ are the charge current density generated in 
the CuMnBi wire, the spin Hall angle of CuMnBi, the spin current density injected into 
CuMnBi, the polarization direction of conduction electron spin, 
and the unit vector along the CuMnBi wire direction, respectively. 
The bottom Py wire is used to estimate the spin diffusion length $\lambda_{\rm CuMnBi}$ 
or spin relaxation time $\tau_{\rm CuMnBi}$ 
(by using $\lambda_{\rm CuMnBi} =\sqrt{D\tau_{\rm CuMnBi}}$ where $D$ is the diffusion constant) 
of CuMnBi, as explained in the next subsection.

In Fig.~\ref{fig:figure3}, we show ISHE resistances 
$R_{\rm ISHE}$ of ${\rm Cu}_{99.5-x}{\rm Mn}_{x}{\rm Bi}_{0.5}$ 
with two different Mn concentrations ($x = 4.2$ and $10.6$). 
$R_{\rm ISHE}$ is defined as the detected voltage drop $V$ along the CuMnBi wire 
(proportional to ${\bm j}_{\rm c}$ in Eq.~(\ref{eq1})) divided by 
the injection current $I$ from Py to Cu [see the inset of Fig.~\ref{fig:figure3}(a)]. 
When $B > 0.3$~T (or $< -0.3$~T), $R_{\rm ISHE}$ is fully saturated. 
At 100 K higher than $T_{\rm f} (= 25 \sim 45~K)$ of bulk CuMnBi, 
a clear negative ISHE signal 
[$2\Delta R_{\rm ISHE} \equiv R_{\rm ISHE}(B > 0.3~{\rm T}) - R_{\rm ISHE}(B < -0.3~{\rm T})$] 
is observed for both Mn concentrations. 
This shows that the ISHE occurs at the Bi impurity sites and 
the Mn impurities do not contribute to the ISHE. However, at low temperatures, 
the ISHE signals become smaller and vanishes at around 20 K for Mn 10.6\%.

\begin{figure}
\begin{center}
\includegraphics[width=5.5cm]{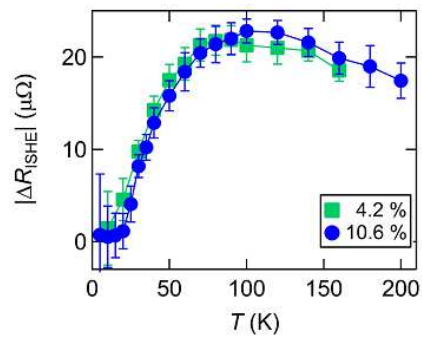}
\caption{Temperature dependence of the amplitudes of ISHE resistances ($|\Delta R_{\rm ISHE}|$) of ${\rm Cu}_{95.3}{\rm Mn}_{4.2}{\rm Bi}_{0.5}$ and ${\rm Cu}_{88.9}{\rm Mn}_{10.6}{\rm Bi}_{0.5}$.}
\label{fig:figure4}
\end{center}
\end{figure}

We show the temperature dependence of the amplitudes of ISHE resistances
$|\Delta R_{\rm ISHE}|$ for ${\rm Cu}_{99.5-x}{\rm Mn}_{x}{\rm Bi}_{0.5}$ 
with two different Mn concentrations ($x = 4.2$ and $10.6$) in Fig.~\ref{fig:figure4}.
With decreasing temperature from $T = 200$~K, 
$|\Delta R_{\rm ISHE}|$ increases because the spin diffusion length of the Cu bridge becomes longer. 
However, it starts to decrease at $80\sim120$~K and vanishes at $10\sim20$~K. 

\begin{figure}
\begin{center}
\includegraphics[width=5.5cm]{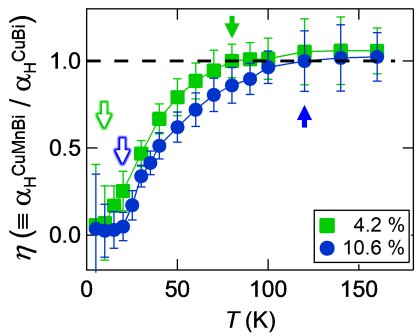}
\caption{Temperature dependence of $\eta$, the spin Hall angle of 
${\rm Cu}_{99.5-x}{\rm Mn}_{x}{\rm Bi}_{0.5}$ ($x= 4.2$ and 10.6) normalized by 
that of ${\rm Cu}_{99.5}{\rm Bi}_{0.5}$. The solid and open arrows 
indicate $T^{*}$ and $T_{\rm f}$, respectively.}
\label{fig:figure5}
\end{center}
\end{figure}

In order to quantitatively evaluate the reduction of the ISHE and the vanishing ISHE 
in CuMnBi, we obtained the spin Hall angle of ${\rm Cu}_{99.5-x}{\rm Mn}_{x}{\rm Bi}_{0.5}$ normalized by 
that of ${\rm Cu}_{99.5}{\rm Bi}_{0.5}$, i.e., 
$\displaystyle \eta \equiv \frac{\alpha_{\rm H}^{\rm CuMnBi}}{\alpha_{\rm H}^{\rm CuBi}}$, 
as a function of temperature~\cite{Niimi_PRL_2015}. 
We note that the ISHE in CuMnBi originates from the skew scattering 
at the Bi impurity sites~\cite{niimi_2012,fert_levy_2011,Niimi_PRL_2015,niimi_prb_2014}. 
In other words, $\alpha_{\rm H}^{\rm CuMnBi}$ is independent of the Mn concentration. 
Therefore, in principle, $\eta$ should be 1. This is realized 
in the high temperature region for all the Mn concentrations, as shown in Fig.~\ref{fig:figure5} 
and Fig.~S2 in Ref.~\onlinecite{supplemental_material}. 
With decreasing temperature, on the other hand, 
$\eta$ starts to deviate from 1 at $T^{*}$ and becomes smaller 
with decreasing temperature. 
Here $T^{*}$ is defined as the temperature where 
$\displaystyle \eta(T+\Delta T) -\eta(T)$ becomes smaller than 4\% 
(in the present case, $\Delta T \sim 10$~K) as we increase $T$. 
The temperature dependence of $\eta$ can be explained as follows. 
Above $T^{*}$, the magnetic moments at the Mn sites fluctuate very fast, as shown in Fig.~\ref{fig:figure1}(a). 
The conduction electron spins are not affected by the fluctuation and 
scattered at the Bi sites keeping the condition of Eq.~(1). 
This corresponds to the PM state.
Below $T^{*}$, however, the magnetic moments at the Mn sites start 
to couple with the conduction electron spins. This coupling 
induces depolarization of the conduction electron spins, as illustrated in Fig.~\ref{fig:figure1}(b). 
These facts can be explained by the simple equation:
\begin{eqnarray*}
\alpha_{\rm H}^{\rm CuMnBi} = \alpha_{\rm H}^{\rm CuBi} \sin\theta
\end{eqnarray*}
where $\theta$ is the angle between the spin polarization direction $z$ and 
the spin current direction $y$. 
The injected conduction spins are polarized along the $z$ axis and thus
$\theta$ is originally set to $\pi/2$ [see Fig.~\ref{fig:figure1}(a)], 
but the spin polarization direction of conduction electron 
is randomized by the coupling with the Mn moments below $T^{*}$ [see Fig.~\ref{fig:figure1}(b)]. 
This coupling induces a reduction of $\theta$, and thus
$\eta$, with decreasing temperature. 
A \textit{partial} reduction of $\eta$ was 
already reported in the previous work~\cite{Niimi_PRL_2015}, but 
the main point of the present work is that $\eta$ completely vanishes at low temperatures, 
as shown in the open arrows in Fig.~\ref{fig:figure5}. 
This situation corresponds to Fig.~\ref{fig:figure1}(c). 
Below $T_{\rm f}$, the magnetic moments of Mn impurities starts to freeze in random directions. 
In such a case, the conduction electron spins are affected 
by the random directions of the Mn moments and the averaged 
$\theta$ value becomes zero. 
Therefore, it is natural to definite the temperature where $\eta(T) = 0$ as $T_{\rm f}$. 
More quantitatively, in the same way as we have done for $T^{*}$, 
$T_{\rm f}$ has been defined as the temperature where 
$\displaystyle \eta(T+\Delta T) -\eta(T)$ becomes larger than 4\% 
(in the present case, $\Delta T = 5$~K) in the low temperature region. 
We will discuss the validity of this definition in more detail in Sec.~IV~A.

\subsection{NLSV measurements with CuMnBi nanowire}

\begin{figure}[htbp]
\begin{center}
\includegraphics[width=5.5cm]{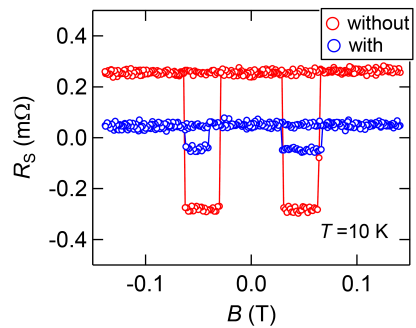}
\caption{NLSV signals with and without the Cu$_{95.3}$Mn$_{4.2}$Bi$_{0.5}$ 
wire measured at $T=10$~K.}
\label{fig:figure6}
\end{center}
\end{figure}

\begin{figure}[htbp]
\begin{center}
\includegraphics[width=5.5cm]{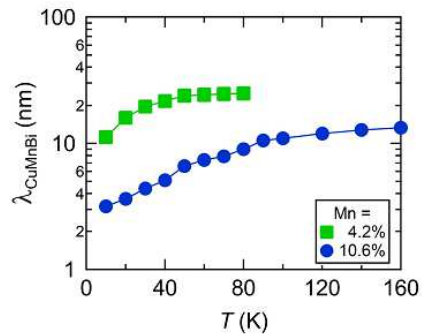}
\caption{Temperature dependence of $\lambda_{\rm CuMnBi}$ for 
${\rm Cu}_{99.5-x}{\rm Mn}_{x}{\rm Bi}_{0.5}$ ($x= 4.2$ and 10.6).}
\label{fig:figure7}
\end{center}
\end{figure}

In order to evaluate the spin diffusion length $\lambda_{\rm CuMnBi}$ and the spin diffusion time 
$\tau_{\rm CuMnBi}$ of CuMnBi, we performed NLSV measurements 
with and without the CuMnBi nanowire~\cite{Niimi_PRL_2015,Niimi_rpp_2015}. 
The typical NLSV data are shown in Fig.~\ref{fig:figure6}. 
By inserting the CuMnBi wire, the NLSV signal $R_{\rm S}^{\rm with}$ detected at 
the bottom Py wire in the inset of Fig.~\ref{fig:figure3}(a) is reduced, 
compared to the nonlocal spin valve signal 
without the CuMnBi wire $R_{\rm S}^{\rm without}$.
This is because most of the spin current flowing in the Cu channel shown in 
the inset of Fig.~\ref{fig:figure3}(a) is absorbed into the CuMnBi wire and 
the rest of the spin current reaches the bottom Py wire, leading to a 
reduced nonlocal spin signal. 
From the reduction rate $R_{\rm S}^{\rm with}/R_{\rm S}^{\rm without}$, 
the spin diffusion length $\lambda_{\rm CuMnBi}$ can be obtained, 
as depicted in Fig.~\ref{fig:figure7}. 
In the present case, $\lambda_{\rm CuMnBi}$ is comparable to or smaller than 
the thickness of the CuMnBi nanowire (20~nm). 
Thus, we have adopted the one-dimensional spin diffusion model~\cite{Niimi_PRL_2015,Niimi_rpp_2015}. 

\section{Discussions}

\subsection{Evaluation of $T_{\rm f}$ of spin glass nanowire}

\begin{figure}
\begin{center}
\includegraphics[width=7cm]{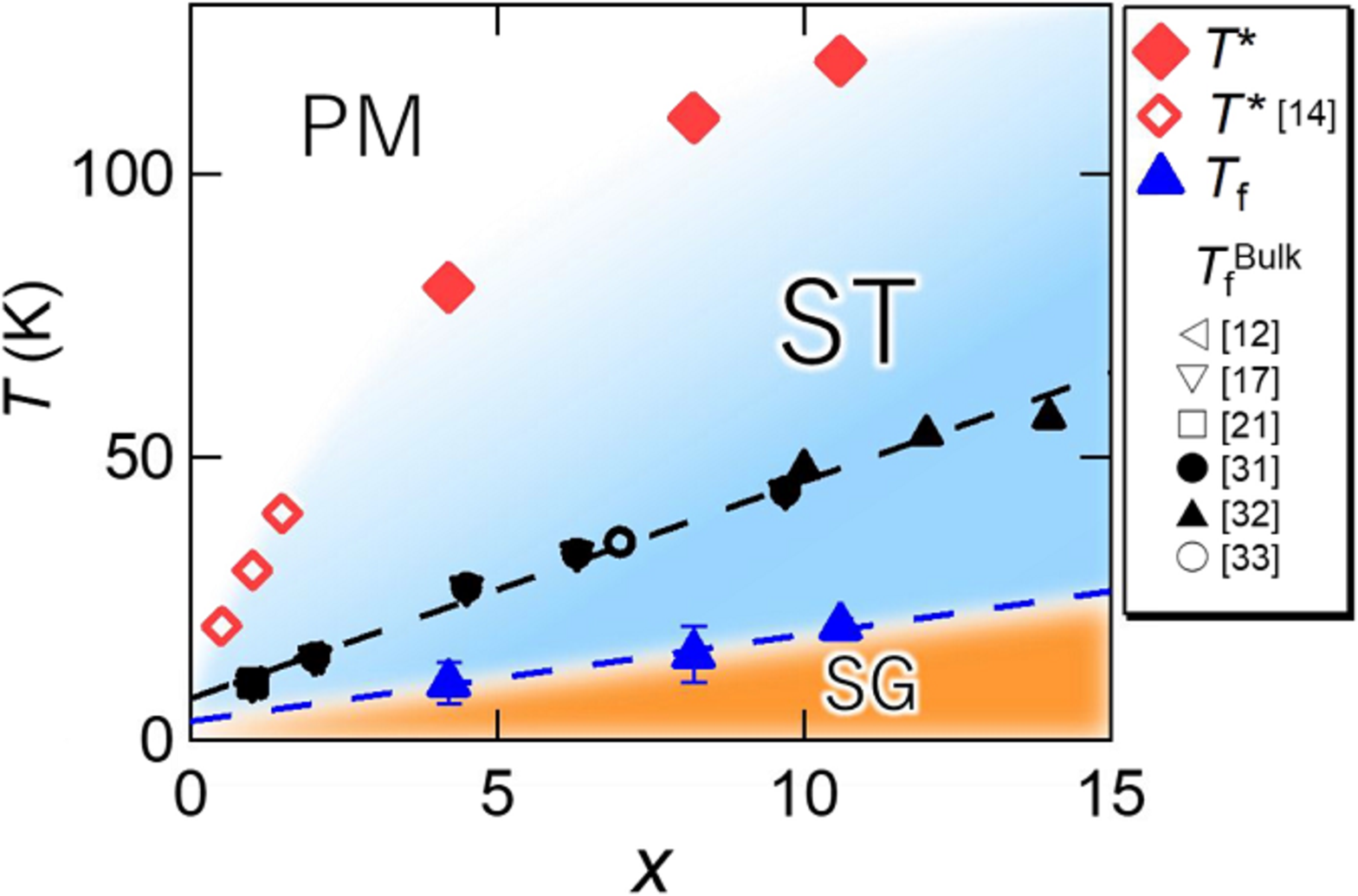}
\caption{Phase diagram of SG proposed in the present work. 
$T_{\rm f}$ and $T^{*}$ of CuMnBi nanowires are indicated by 
the solid blue triangle and the solid red square, respectively. 
The data for ${T_{\rm f}}^{\rm Bulk}$, and  for $T^{*}$ at lower  values of $x$ than our samples,  are 
taken from Refs.~\onlinecite{Nagata_1979,Uemura_PRB_1985,Alloul_PRL_1979,Ford_PRB_1976,Gibbs_1985,Kenning_PRL_1987} 
and Ref.~\onlinecite{Niimi_PRL_2015}, respectively.}
\label{fig:figure8}
\end{center}
\end{figure}

As mentioned in Sec.~III~B, 
we argue that the temperature where $\eta$ vanishes corresponds to $T_{\rm f}$ of 
SG nanowire, below which the SG state starts to emerge. 
To check the validity of the determination of $T_{\rm f}$ in SG nanowires using the ISHE measurements, 
we plot $T_{\rm f}$ as well as $T^{*}$ as a function of the Mn concentration $x$ 
in Fig.~\ref{fig:figure8}.
It is obvious that $T_{\rm f}$ determined from the spin transport measurements 
increases linearly with $x$, as in the case of bulk SG~\cite{Nagata_1979,Uemura_PRB_1985,Alloul_PRL_1979,Ford_PRB_1976,Gibbs_1985,Kenning_PRL_1987}. 
Note that a linear fit  passes  through the origin neither  for the nanowire nor bulk SGs. 
Moreover, $T_{\rm f}$ of our nanowire 
is about half of that of bulk CuMnBi determined from 
the magnetization measurements [see Figs.~\ref{fig:figure2} and \ref{fig:figure8}]. 
A similar size effect was already established 
for thin SG films~\cite{Kenning_PRL_1987,Kenning_PRB_1990}. 
From these experimental facts, we conclude that 
$T_{\rm f}$ of SG nanowires can be determined with the ISHE measurements 
and the ISHE measurements are powerful for the detection of 
tiny magnetic signals via spin current on-chip~\cite{wei_nat_commun_2012}.

On the other hand, to fully prove that $T_{\rm f}$ obtained with the ISHE measurements 
correspond to those measured with standard magnetometry, 
one needs micro-SQUID measurements. That would be the future work.

\subsection{Spin treacle in spin glass}

Now we focus on the intermediate regime ($T_{\rm f} < T < T^{*}$) 
in between PM and SG, as shown in the light-blue region in Fig.~\ref{fig:figure8}. 
In general, $T_{\rm f}$ can be determined from a cusp 
in the magnetization versus temperature curve. 
It is believed that the SG and PM states exist 
below and above $T_{\rm f}$, respectively. 
However, the present spin transport measurements clearly show that 
there is another regime in between the two, which we name ``spin treacle (ST)".
Even for $T > T_{\rm f}$, 
the conduction electron spins feel magnetic fluctuations of the magnetic moments, 
as illustrated in Fig.~\ref{fig:figure1}(b). 
Such a subtle fluctuation can be detected by spin current. 
In fact, Campbell \textit{et al}.~\cite{Campbell_PRL_1994} pointed out 
the possibility of a regime of stretched exponential local spin correlations 
just above $T_{\rm f}$ in AgMn and AuFe alloys 
using $\mu$SR measurements 
and concluded that this was an intrinsic precursor to SG freezing. 
The present work not only reinforces the appearance of ST but also 
extends the applicable scope to the long-wavelength spin current rather than 
the local response seen by muons. 
In addition, in Ref.~\onlinecite{Campbell_PRL_1994} the concentration dependence of $T^{*}$ 
was not shown in detail.
Here we see that $T_{\rm f}$ linearly increases with increasing $x$, while 
$T^{*}$ seems to be proportional to $x^{\gamma}$ where $\gamma \approx 1/3$, 
although further experimental and theoretical works are desirable 
to fully understand this exponent.

We examine the ST in another SG system. 
For this purpose, we have performed ISHE measurements in AuFe, another typical SG alloy. 
As shown in Fig.~S3 in Ref.~\onlinecite{supplemental_material}, 
$\Delta R_{\rm ISHE}$ starts to decrease at 
$T^{*} \approx 5T_{\rm f}$ and becomes flat at $T_{\rm f}$. 
This result clearly shows that ST is a common feature in SG materials. 
The difference between CuMnBi and AuFe is that 
$\Delta R_{\rm ISHE}$ vanishes or takes a finite value below $T_{\rm f}$. 
This originates from the spin-orbit interaction of host metal. 
Cu has so weak spin-orbit interaction that the ISHE cannot be detected. 
Thus, a small number of Bi impurities are added to induce ISHE~\cite{niimi_2012}. 
In the present case, the number of Mn impurities are more than several times larger 
than that of Bi impurities. 
The injected spin current loses the information of spin direction due to 
the slowing dynamics of Mn impurities before 
being scattered at the Bi site, leading to zero ISHE signal. 
On the other hand, Au has an intrinsic ISHE due to its stronger spin-orbit interaction~\cite{niimi_prb_2014}. 
The injected spin current can be skew scattered by Au atoms (even without being scattered by Fe impurities), 
leading to a finite ISHE signal. 
The difference of the spin-orbit interaction also appears 
in anomalous Hall effect (AHE)~\cite{taniguchi_prl_2004,campbell_epl_2004,nagaosa_rmp_2010}:
the AHE in AuFe shows a typical cusp structure at $T_{\rm f}$ 
below which the difference between ZFC and FC in the AHE can be seen, while
such a clear cusp cannot be seen in the AHE in CuMn and CuMnBi. 

Compared to magnetization measurements, a larger magnetic field is needed to 
clearly detect the AHE and the ISHE (at least 0.3~T for the ISHE in the present setup). 
Such a comparably large magnetic field would slightly reduce $T_{\rm f}$~\cite{taniguchi_prl_2004,campbell_epl_2004}. 
To avoid the possible reduction of $T_{\rm f}$ with an applied magnetic field, 
the shape of a spin injection ferromagnetic nanowire in the inset of Fig.~\ref{fig:figure2}(a)
can be changed so as to have the spin polarization along the Cu bridge 
without the magnetic field~\cite{kimura_prl_2007}. 
As for AuFe, a Hall cross device consisting of a ferromagnet with perpendicular magnetic anisotropy 
and a AuFe cross bar can also be used for the determination of $T_{\rm f}$ 
in the zero magnetic field limit~\cite{seki,sugai}.

\subsection{Evaluation of correlation time of localized moments in spin treacle}

\begin{figure}
\begin{center}
\includegraphics[width=8.5cm]{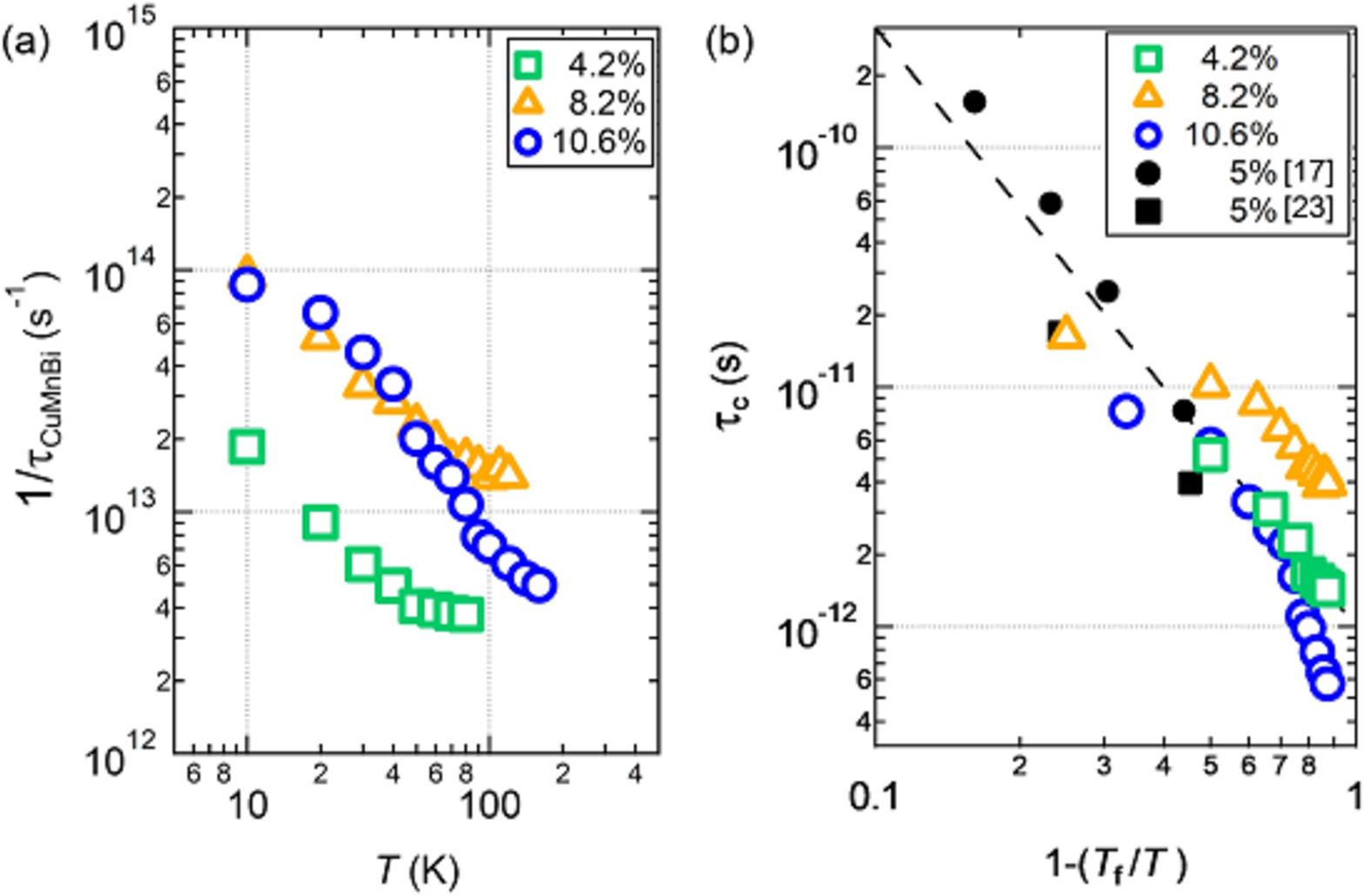}
\caption{(a) Temperature dependence of the spin relaxation rate $1/\tau_{\rm CuMnBi}$ of 
${\rm Cu}_{99.5-x}{\rm Mn}_{x}{\rm Bi}_{0.5}$ ($x= 4.2$, 8.2 and 10.6). 
(b) Correlation time $\tau_{\rm c}$ of Mn moments above $T_{\rm f}$
determined by spin relaxation time measurements with $\Delta \approx 4k_{\rm B}T_{\rm f}$ 
in Eq.~(\ref{eq2}). $\tau_{\rm c}$ determined 
by zero-field $\mu$SR~\cite{Uemura_PRB_1985} and neutron spin echo measurements~\cite{mezei_jmmm_1979} for samples of similar concentrations of Mn, but without Bi, 
are superimposed onto the same graph. The broken line shows a 
power law $(1-T_{\rm f}/T)^{-b}$ with $b \approx 2.5$ [see Eq.~(\ref{eq3})].}
\label{fig:figure9}
\end{center}
\end{figure}

We have further investigated the correlation time $\tau_{\rm c}$ of localized magnetic moments 
in ST. For this purpose, we first obtained the spin relaxation time 
$\tau_{\rm CuMnBi}$ of conduction electrons from Fig.~\ref{fig:figure7} using the relation 
$\lambda_{\rm CuMnBi} =\sqrt{D\tau_{\rm CuMnBi}}$. 
The inverse of $\tau_{\rm CuMnBi}$, i.e., spin relaxation rate, is plotted as a function of $T$ 
in Fig.~\ref{fig:figure9}(a). With decreasing temperature, 
the spin relaxation rate increases and tends to be saturated 
especially for $x=10.6$\% 
as $T$ approaches $T_{\rm f}$. 
A similar temperature dependence has been discussed 
in the muon depolarization rate~\cite{Uemura_PRB_1985,Campbell_PRL_1994} and also
in the linewidth of ESR spectrum~\cite{date_jpsj_1969,chien_jap_1994}. 
The $\mu$SR and spin transport measurements 
detect the precession of muon and conduction electron spins, respectively. 
At high enough temperatures, the muon and conduction electron spins 
do not feel the fast motion of Mn localized moments. 
With decreasing temperature, the fluctuation of the localized moments 
becomes slower, and the muon and conduction 
electron spins start to couple with this fluctuation, resulting in the enhancement 
of $1/\tau_{\rm CuMnBi}$.

Now we relate $1/\tau_{\rm CuMnBi}$ with $\tau_{\rm c}$ in SG nanowire. 
As with muon relaxation~\cite{Uemura_PRB_1985} and 
inelastic spin-flip neutron scattering~\cite{mezei_jmmm_1979}, $\tau_{\rm c}$ can be deduced
from the spin current relaxation, as we shall discuss below.
We start with the Kubo-Toyabe theory commonly used 
in the analysis of $\mu$SR. In the present case, however, the ``spin" is 
the polarization of the spin current. 
In the high frequency limit (i.e., the motionally-narrowed regime), 
there is an explicit form for the relaxation time 
(see Ref.~\onlinecite{Uemura_PRB_1985} and references therein).
On the other hand, if we go to lower frequencies in the model, i.e., close to the freezing temperature, 
the Kubo-Toyabe approach that assumes a single effective field with a decay of a simple exponential 
is not sufficient for the spin current. 
In Ref.~\onlinecite{Niimi_PRL_2015}, a quantity  
$1/(T_{1})_{\rm Mn}$ was defined from the asymptotic form of the longitudinal correlation function 
for the spin polarization of the spin current 
$\displaystyle G_{zz}(t) \propto \exp \left[ - t/(T_{1})_{\rm Mn} \right]$ 
(see Ref.~\onlinecite{Niimi_PRL_2015} and references therein).
$(T_{1})_{\rm Mn}$ denotes the relaxation time of the spin current affected 
only by the fluctuation of Mn moments. In the high frequency limit,
we have the following relation between $(T_{1})_{\rm Mn}$ and $\tau_{\rm c}$: 
\begin{eqnarray*}
\tau_{\rm c} &=& \frac{h^{2}}{2{\Delta}^2} \frac{1}{(T_{1})_{\rm Mn}}, 
\end{eqnarray*}
where $h$ and $\Delta$ are the Planck constant, 
the magnitude of the effective field acting on the spin current through the exchange 
interactions with the Mn moments, respectively. 
We note that the effective field acting on the spin current 
is a combination at long wavelengths, in contrast to that acting on the muon spin, 
which is {\it local}.
The spin current polarization relaxes by two mechanisms, i.e.,
the exchange with the Mn moments and spin-orbit interactions at the Bi sites, as follows:
\begin{eqnarray*}
\frac{1}{\tau_{\rm CuMnBi} } =  \frac{1}{(T_1)_{\rm Mn}} +\frac{1}{\tau_{\rm CuBi}}.
\end{eqnarray*}
Therefore, $\tau_{\rm c}$ can be given by
\begin{eqnarray}
\tau_{\rm c}  = 
\frac{h^{2}}{2{\Delta}^2}\left( \frac{1}{\tau_{\rm CuMnBi}} - \frac{1}{\tau_{\rm CuBi}} \right). 
\label{eq2} 
\end{eqnarray}

In order to deduce $\tau_{\rm c}$ as a function of temperature for a given concentration, 
we need a value for $\Delta$. 
From the microscopic viewpoint, the field acting on the spin current, which is a macroscopic quantity, 
is generated by a sum of operators on the Mn moments and should scale as the concentration, 
similarly to the molecular field leading to cooperative magnetic order. 
We therefore take $\Delta=a k_{\rm B}T_{\rm f}$, i.e., proportional to 
the freezing temperature $T_{\rm f}$ 
with a prefactor $a$ that should be considerably larger than 1, 
to reflect the frustration of the SG interactions: 
the freezing temperature is expected to be an order of magnitude smaller than the molecular fields.

According to the results of the $\mu$SR in bulk SGs~\cite{Uemura_PRB_1985}, 
$\tau_{\rm c}$ obeys the following temperature law:
\begin{eqnarray}
\tau_{\rm c} \propto \left( 1-\frac{T_{\rm f}}{T} \right)^{-b},
\label{eq3}
\end{eqnarray}
where $b$ is the exponent of $1-T_{\rm f}/T$.
Thus, we plot $\tau_{\rm c}$ with $a \approx 4$ as a function of $1-T_{\rm f}/T$ 
in Fig.~\ref{fig:figure9}(b). 
It nicely follows a power law of $(1-T_{\rm f}/T)^{-b}$ with $b \approx 2.5$. 
In addition, the data obtained with different experimental methods 
($\mu$SR~\cite{Uemura_PRB_1985} and neutron scattering~\cite{mezei_jmmm_1979}) 
for bulk samples of CuMn are superimposed onto the same fitting line. 
The result gives consistent estimates of the spin-spin correlation time for Mn, 
despite the fact that the length scales probed are very different: 
while both muons and neutrons are sensitive to local dipole interactions, 
the spin current can detect the spin fluctuation over long wavelengths. 
This fact clearly verifies that spin relaxation measurements are comparable techniques to 
$\mu$SR and neutron scattering measurements but more suitable for nanoscale 
frustrated magnetic systems to detect the slowing magnetic dynamics.

Finally, we mention the differences of the results and analyses between 
the present work and the previous publication~\cite{Niimi_PRL_2015}, as well as 
the differences between spin current and muons as magnetic probes. 
In Ref.~\onlinecite{Niimi_PRL_2015}, the CuMnBi SG nanowires with lower concentrations of Mn led to 
lower freezing temperatures $T_{\rm f}$ and it was not then clear whether the spin Hall angle was 
reduced by a finite amount or would vanish completely. 
The higher concentrations studied here allow us to see clearly a vanishing spin Hall angle. 
In other words, the spin current is completely depolarized at $T_{\rm f}$. 
The interpretation of Ref.~\onlinecite{Niimi_PRL_2015} was also based on the Kubo-Toyabe model 
but the characteristic frequency was assumed to be 
$\displaystyle \tau_{\rm c} \propto \left( 1-\frac{T_{\rm f}}{T} \right)^{-2}$, 
while in the present work, we have experimentally 
determined $\displaystyle \tau_{\rm c} \propto \left( 1-\frac{T_{\rm f}}{T} \right)^{-b}$ 
with $b \approx 2.5$. 
We note that the exponent $b$ is expressed, 
in terms of a conventional phase transition, 
as the combination of $b=z\nu$, with the dynamical critical exponent $z$ and 
the power $\nu$ relating the time scale to a correlation length that diverges as 
$\left( T-T_{\rm f} \right)^{-\nu}$. 
In the case of the spin glass transition, this correlation length is of 
a four-spin correlation function~\cite{Binder_Young_1986}. 
The previous theory could ``qualitatively'' reproduce 
the reduction of the spin Hall angle far above $T_{\rm f}$. 
However, from comparison with the present experimental results 
where $T_{\rm f}$ is higher, it has turned out that the previous theory could not explain 
the vanishing spin Hall angle near $T_{\rm f}$.
From the muon experiments~\cite{Uemura_PRB_1985,Campbell_PRL_1994}, 
we may expect the assumption of a simple exponential decay 
to be invalid in the ST regime: the muon spin polarization shows ``stretched'' exponential 
correlations, $\exp \left[ -(\lambda t)^{\beta} \right]$ 
with $\beta < 1$ in this regime, as detailed in Ref.~\onlinecite{Campbell_PRL_1994}. 
The limitation for the direct comparison with spin current is 
that in $\mu$SR experiments, a fully polarized muon spin is 
injected into the SG material and is eventually trapped into specific sites of the SG material 
where it precesses around a local random field that is frozen below $T_{\rm f}$. 
This leads to a finite average polarization because the muon spin does not lose the component 
of its polarization projected along the direction of the particular frozen field around 
which it precesses. On the other hand, in spin transport measurements, 
the spin current travels diffusively in the SG material. 
The information of spin angular momentum is kept only in $\tau_{\rm CuMnBi}$. 
Such a diffusive motion of spin current 
was not taken into account in the theory of Ref.~\onlinecite{Niimi_PRL_2015}.
Inclusion of these last two points is not easy at all, 
but highly desirable to fully understand the spin current dynamics in frustrated spin systems.

\section{Conclusions}

In conclusion, the spin transport measurements demonstrate 
that a ST regime emerges between the SG and PM phases. 
The spin Hall angle of CuMnBi is constant in the temperature 
range of $T > T^{*}$ corresponding to the PM state, but starts to decrease below 
$T^{*}$ and eventually vanishes. 
$T_{\rm f}$ determined from the spin Hall angle measurements linearly increases with 
increasing the magnetic impurity concentration, as in the case of bulk SG. 
In the temperature region of $T_{\rm f} < T < T^{*}$, we find the ST regime, 
which would correspond to the regime hypothesized by $\mu$SR experiments 
but has not been distinguished by other conventional experimental 
techniques for bulk SG materials. 
Furthermore, the slowing dynamics of localized spins in the ST regime 
can also be quantitatively evaluated 
through the spin relaxation measurements of conduction electrons.
The present result not only demonstrates how quantitative characterization of 
magnetic fluctuations on nanometer-scale samples is possible using spin transport measurements, 
but also paves the way to study other magnetic systems where the spin fluctuations are essential.

\acknowledgments 
We acknowledge fruitful discussions with H. Kawamura and K. Aoyama. 
This work was supported by JSPS KAKENHI 
(Grant Nos. JP16H05964, JP16H04023, JP17K18756, JP17H02927, 
JP19K21850, JP19H00656, JP19H05826, JP26103002, and JP26103005), 
JST-ERATO (Grant No. JPMJER1402), JST-CREST (Grant No. JPMJCR19J4), 
NSFC (Grant No. Y81Z01A1A9), CAS (Grant No. Y929013EA2), 
UCAS (Grant No. 110200M208), 
BNSF (Grant No. Z190011), 
the Mazda Foundation, Shimadzu Science Foundation, 
Yazaki Memorial Foundation for Science and Technology, 
SCAT Foundation, and the Murata Science Foundation.

\end{document}